\documentclass[12pt]{article}
\usepackage{graphicx}
\usepackage{amsmath}
\usepackage{amssymb}
\usepackage{caption2}
\begin{document}
\bibliographystyle{prsty}
\begin{center}
{\large {\bf \sc{   Analysis  of the  vertices $DDV$ and $D^*DV$ with light-cone QCD sum rules  }}} \\[2mm]
Zhi-Gang Wang \footnote{E-mail,wangzgyiti@yahoo.com.cn.  }    \\
Department of Physics, North China Electric Power University,
Baoding 071003, P. R. China
\end{center}

\begin{abstract}
In this article, we study the vertices $DDV$ and $D^*DV$ with
  the light-cone QCD sum rules.  The  strong
coupling constants $g_{DDV}$ and $f_{D^*DV}$ play an important role
in understanding the final-state re-scattering effects  in the
hadronic $B$ decays. They are related to the basic parameters
$\beta$ and $\lambda$ respectively in the heavy quark effective
Lagrangian, our numerical values are smaller than the existing
estimations.
\end{abstract}


\section{Introduction}

Final-state interactions (or re-scattering effects) play an
important role in the hadronic $B$ decays \cite{fsi,CHHY}. However,
it is very difficult to take them into account  in a systematic way
due  to the nonperturbative nature of the multi-particle dynamics.
In practical calculations, we can resort to phenomenological models
to outcome the difficult. The one-particle-exchange model is typical
(for example, see Ref.\cite{CHHY}), in this picture, the soft
re-scattering  of the intermediate states in  two-body channels with
one-particle exchange makes the main contributions. The
phenomenological Lagrangian contains many input parameters, which
describe the strong couplings  among the charmed mesons in the
hadronic $B$ decays.

In the following, we write down the relevant  phenomenological
Lagrangian, which describes the strong interactions of the $DDV$ and
$D^*DV$ \cite{CHHY},
\begin{eqnarray}
\mathcal {L}&=&ig_{DDV}D_i
(\overrightarrow{\partial_\mu}-\overleftarrow{\partial_\mu})D_j
V^\mu_{ij} \nonumber \\
&&+2f_{D^*DV}\epsilon_{\mu\nu\alpha\beta}\partial^\mu V^{\nu}_{ij}
\left[D_i\left(\overrightarrow{\partial^\alpha}-\overleftarrow{\partial^\alpha}\right)D^{*\beta}_j
-D^{*\beta}_i\left(\overrightarrow{\partial^\alpha}-\overleftarrow{\partial^\alpha}\right)D_j
\right] \, ,\nonumber\\
D&=&(D^0,D^+,D_s) \, , \nonumber\\
D^*&=&(D^{*0},D^{*+},D^*_s) \, , \nonumber\\
V&=&\left(\begin{array}{ccc}
\frac{\rho^{0}}{\sqrt{2}}+\frac{\omega}{\sqrt{2}}&\rho^{+}&K^{*+}\\
\rho^{-}&-\frac{\rho^{0}}{\sqrt{2}}+\frac{\omega}{\sqrt{2}}&
K^{*0}\\
K^{*-} &\bar{K}^{*0}&\phi
\end{array}\right) \, .
\end{eqnarray}

 The strong coupling constants  $g_{DDV}$ and $f_{D^*DV}$  in the
phenomenological Lagrangian can be related to the basic parameters
$\beta$ and $\lambda$ in the heavy quark effective Lagrangian (one
can consult Ref.\cite{HQEFT} for the heavy  quark effective
Lagrangian and relevant parameters.\footnote{\begin{eqnarray} {\cal
L}&=& i \langle H_b v^\mu D_{\mu}^{ba} {\bar H}_a \rangle + i\beta
\langle H_bv^\mu\left({\cal V}_\mu-\rho_\mu\right)_{ba}{\bar
H}_a\rangle + i \lambda \langle H_b \sigma^{\mu\nu}
F_{\mu\nu}(\rho)_{ba} {\bar H}_a\rangle~~. \nonumber
\end{eqnarray} } ),

\begin{eqnarray}
g_{DDV}&=& \frac{\beta g_V}{\sqrt{2}}\, , \nonumber \\
 f_{D^*DV}&=&\frac{\lambda g_V}{\sqrt{2}}\, ,
\end{eqnarray}
where  $g_V=5.8$ from the vector meson dominance theory
\cite{VMDgV}.

 In this article, we study the strong coupling constants
 $g_{DDV}$ and $f_{D^*DV}$ with the light-cone QCD
 sum rules \cite{LCSR,LCSRreview}. The strong coupling constants $g_{BB\rho}$, $g_{DD\rho}$,
 $f_{B^*B\rho}$ and $f_{D^*D\rho}$ have been
 calculated with the light-cone QCD sum rules in Ref.\cite{LiHuang}, I
failed  to take notice of that work at beginning.

   The light-cone QCD sum rules  carry
out   operator product expansion near the light-cone, $x^2\approx
0$, instead of  short distance, $x\approx 0$, while the
nonperturbative matrix elements are parameterized by the light-cone
distribution amplitudes (which are classified according to their
twists)  instead of
 the vacuum condensates \cite{LCSR,LCSRreview}.
 The nonperturbative
 parameters in the light-cone distribution amplitudes are calculated by   the conventional QCD  sum rules
 and the  values are universal \cite{SVZ79}.

The article is arranged as: in Section 2, we derive the strong
coupling constants  $g_{DDV}$ and  $f_{D^*DV}$ with  the light-cone
QCD sum rules; in Section 3, the numerical result and discussion;
and in Section 4, conclusion.

\section{Strong coupling constants  $g_{DDV}$ and  $f_{D^*DV}$ with light-cone QCD sum rules}

We study the strong coupling constants  $g_{DDV}$ and $f_{D^*DV}$
 with the two-point correlation functions $\Pi_{ij}(p,q)$ and $\Pi^{ij}_{\mu}(p,q)$,
\begin{eqnarray}
\Pi_{ij}(p,q)&=&i \int d^4x \, e^{-i q \cdot x} \,
\langle 0 |T\left\{J_i(0) J_j^+(x)\right\}|V_{ij}(p)\rangle \, , \\
\Pi^{ij}_{\mu}(p,q)&=&i \int d^4x \, e^{-i q \cdot x} \,
\langle 0 |T\left\{J^i_\mu(0) J_j^+(x)\right\}|V_{ij}(p)\rangle \, , \\
J_i(x)&=&{\bar q}_i(x)i\gamma_5   c(x)\, ,  \nonumber \\
J^i_\mu(x)&=&{\bar q}_i(x)\gamma_\mu   c(x)\, ,
\end{eqnarray}
where the currents $J_i(x)$ interpolate the pseudoscalar mesons
$D^0$, $D^+$, $D_s$ and the currents $J^i_\mu(x)$ interpolate the
vector  mesons $D^{*0}$, $D^{*+}$, $D^*_s$. The external states
$\rho$, $K^*$ and $\phi$ have the four momentum $p_\mu$ with
$p^2=m_\rho^2$, $m_{K^*}^2$ and $m_{\phi}^2$, respectively.

According to the basic assumption of current-hadron duality in the
QCD sum rules \cite{SVZ79}, we can insert  a complete series of
intermediate states with the same quantum numbers as the current
operators $J_i(x)$ and $J^i_{\mu}(x)$ into the correlation functions
$\Pi_{ij}(p,q)$ and $\Pi^{ij}_{\mu}(p,q)$ to obtain the hadronic
representation. After isolating the ground state contributions from
the pole terms of the mesons $D_i$ and $D^*_i$, we get the following
results,
\begin{eqnarray}
\Pi_{ij }(p,q) &=&\frac{f_{D_i}f_{D_j}M_{D_i}^2M_{D_j}^2g_{D_iD_j
V_{ij}}}
{(m_i+m_c)(m_j+m_c)\left\{M_{D_i}^2-(q+p)^2\right\}\left\{M_{D_j}^2-q^2\right\}}2\epsilon
\cdot q    + \cdots    \nonumber\\
&=& \Pi^P_{ij}(p,q) \epsilon \cdot q    + \cdots , \\
 \Pi^{ij}_{\mu
}(p,q)&=&\frac{f_{D^*_i}f_{D_j}M_{D_i^*}M_{D_j}^2g_{D^*_iD_j
V_{ij}}}
{(m_j+m_c)\left\{M_{D^*_i}^2-(q+p)^2\right\}\left\{M_{D_j}^2-q^2\right\}}4\epsilon_{\mu\nu\alpha\beta}\epsilon^\nu
p^\alpha q^\beta
    + \cdots    \nonumber \\
    &=& \Pi^V_{ij}(p,q) \epsilon_{\mu\nu\alpha\beta}\epsilon^\nu
p^\alpha q^\beta
    + \cdots ,
\end{eqnarray}
where the following definitions for the weak decay constants have
been used,
\begin{eqnarray}
\langle0 | J_i(0)|D_i(p)\rangle&=&\frac{f_{D_i}M_{D_i}^2}{m_i+m_c}\,, \nonumber\\
\langle0 |
J^i_{\mu}(0)|D^*_{i}(p)\rangle&=&f_{D^*_i}M_{D^*_i}\epsilon_\mu\,.
\end{eqnarray}

 In Eqs.(6-7), we have not shown the contributions from the high
resonances and continuum states explicitly as they are suppressed
due to the double Borel transformation.

In the following, we briefly outline  operator product expansion for
the correlation functions $\Pi_{ij }(p,q)$ and $\Pi_{\mu
}^{ij}(p,q)$   in perturbative QCD theory. The calculations are
performed at  large spacelike momentum regions $(q+p)^2\ll 0$ and
$q^2\ll 0$, which correspond to  small light-cone distance
$x^2\approx 0$ required by  validity of operator product expansion.
We write down the propagator of a massive quark in the external
gluon field in the Fock-Schwinger gauge firstly \cite{Belyaev94},
\begin{eqnarray}
\langle 0 | T \{q_i(x_1)\, \bar{q}_j(x_2)\}| 0 \rangle &=&
 i \int\frac{d^4k}{(2\pi)^4}e^{-ik(x_1-x_2)}
\frac{\not\!k +m}{k^2-m^2} +\cdots\, ,
\end{eqnarray}
where we have neglected the contributions from the gluons $G_{\mu
\nu }$. The contributions proportional to $G_{\mu\nu}$ can give rise
to three-particle (and four-particle) meson distribution amplitudes
with a gluon (or quark-antiquark pair) in addition to the two
valence quarks, their corrections are usually not expected to play
any significant roles\footnote{For examples, in the decay $B \to
\chi_{c0}K$, the factorizable contribution is zero and the
non-factorizable contributions from the soft hadronic matrix
elements are too small to accommodate the experimental data
\cite{WangLH}; the net contributions from the three-valence particle
light-cone distribution amplitudes to the strong coupling constant
$g_{D_{s1}D^*K}$ are rather small, about $20\%$ \cite{Wang0611}. The
contributions of the three-particle (quark-antiquark-gluon)
distribution amplitudes of the mesons are always of minor importance
comparing with the two-particle (quark-antiquark) distribution
amplitudes in the light-cone QCD sum rules.   In our previous work,
we study the four form-factors $f_1(Q^2)$, $f_2(Q^2)$, $g_1(Q^2)$
and $g_2(Q^2)$ of the $\Sigma \to n$ in the framework of the
light-cone QCD sum rules  up to twist-6 three-quark light-cone
distribution amplitudes and obtain satisfactory results
\cite{Wang06}. In the light-cone QCD sum rules,
 we can neglect the contributions from the valence gluons and make relatively rough estimations.}. Substituting the above $c$ quark
propagator and the corresponding $\rho$, $K^*$, $\phi$ mesons
light-cone distribution amplitudes into the correlation functions
$\Pi_{ij}(p,q)$,  $\Pi_{\mu}^{ij}(p,q)$ in Eqs.(3-4) and completing
the integrals over the variables  $x$ and $k$, finally we obtain the
results,
\begin{eqnarray}
\Pi^P_{ij} &=&f_{V_{ij}} m_{V_{ij}} \int_0^1 du
\frac{\phi_\parallel(u)}{AA}+\left[f_{V_{ij}}^\perp-f_{V_{ij}}\frac{m_i+m_j}{m_{V_{ij}}}\right]m_cm_{V_{ij}}^2
\int_0^1 du \frac{h_{||}^{(s)}(u)
}{AA^2} \nonumber\\
&&-\frac{f_{V_{ij}} m_{V_{ij}}^3}{4} \int_0^1 du A(u) \left[
\frac{1}{AA^2}
+\frac{2m_c^2}{AA^3}\right]\nonumber \\
&&-2f_{V_{ij}} m_{V_{ij}}^3 \int_0^1 du \int_0^u d\tau \int_0^\tau
dt C(t)\left[ \frac{1}{AA^2} +\frac{2m_c^2}{AA^3}\right] +\cdots\, ,\\
\Pi^V_{ij} &=&f_{V_{ij}}^\perp  \int_0^1 du
\frac{\phi_\perp(u)}{AA}-\frac{f_{V_{ij}}^\perp m_{V_{ij}}^2}{4}
\int_0^1 du A_\perp(u) \left[ \frac{1}{AA^2}
+\frac{2m_c^2}{AA^3}\right]\nonumber \\
&&+\left[f_{V_{ij}}-f_{V_{ij}}^\perp\frac{m_i+m_j}{m_{V_{ij}}}\right]\frac{m_cm_{V_{ij}}}{2}
\int_0^1 du \frac{g_{\perp}^{(a)}(u) }{AA^2} +\cdots\, ,
\end{eqnarray}

where
\begin{eqnarray}
AA&=&m_c^2-(q+u\,p)^2 \, .\nonumber
\end{eqnarray}
In calculation, the  two-particle vector  mesons light-cone
distribution amplitudes have been used \cite{VMLC}, the explicit
expressions are given in the appendix. The parameters in the
light-cone distribution amplitudes are scale dependent and can be
estimated with the QCD sum rules \cite{VMLC}. In this article, the
energy scale $\mu$ is chosen to be  $\mu=1\rm{GeV}$.

Now we perform the double Borel transformation with respect to  the
variables $Q_1^2=-(p+q)^2$  and  $Q_2^2=-q^2$ for the correlation
functions $\Pi_{ij}^P$ and $\Pi_{ij}^V$ in Eqs.(6-7), and obtain the
analytical expressions of the invariant functions in the hadronic
representation,
\begin{eqnarray}
B_{M_2^2}B_{M_1^2}\Pi^P_{ij}&=&\frac{ 2g_{D_iD_j
V_{ij}}f_{D_i}f_{D_j}M_{D_i}^2
M_{D_j}^2}{(m_c+m_i)(m_c+m_j)M_1^2M_2^2}
\exp\left[-\frac{M^2_{D_i}}{M_1^2}
-\frac{M^2_{D_j}}{M_2^2}\right] +\cdots, \\
B_{M_2^2}B_{M_1^2}\Pi^V_{ij}&=&\frac{ 4f_{D_i^*D_j
V_{ij}}f_{D^*_i}f_{D_j}M_{D^*_i}M_{D_j}^2 }{(m_c+m_j)M_1^2M_2^2}
\exp\left[-\frac{M^2_{D_i^*}}{M_1^2} -\frac{M^2_{D_j}}{M_2^2}\right]
+\cdots,
\end{eqnarray}
where we have not shown  the contributions from the high resonances
and continuum states  explicitly for simplicity.

In order to match the duality regions below the thresholds $s_0$ and
$s_0'$ for the interpolating currents, we can express the
correlation functions $\Pi_{ij}^P$ and $\Pi_{ij}^V$  at the level of
quark-gluon degrees of freedom into the following form,
\begin{eqnarray}
\Pi^{P(V)}_{ij}&=& \int ds \int ds' \frac{\rho_{ij}(s,s')}{
\left\{s-(q+p)^2\right\}\left\{s'-q^2\right\}} \, ,
\end{eqnarray}
where the $\rho_{ij}(s,s')$ are spectral densities, then perform the
double Borel transformation with respect to the variables $Q_1^2$
and $Q_2^2$ directly. However, the analytical expressions of the
spectral densities $\rho_{ij}(s,s')$ are hard to obtain, we have to
resort to some approximations.  As the contributions
 from the higher twist terms  are  suppressed by more powers of
 $\frac{1}{m_c^2-(q+u\,p)^2}$ (or $\frac{1}{M^2}$), the net contributions of the  twist-3 and twist-4
  terms are of minor
importance (also see the sum rules for the strong coupling constants
$G_S$($D_{s0} D_s^* \phi $) and $G_A$($ D_{s1}D_s \phi$) in
Ref.\cite{Wang07}), the continuum subtractions will not affect the
results remarkably. The dominating contributions come from the
two-particle twist-2 terms involving the $\phi_\parallel(u)$ and
$\phi_\perp(u)$. We perform the same trick as
Refs.\cite{Belyaev94,Kim} and expand the amplitudes
$\phi_\parallel(u)$ and $\phi_\perp(u)$ in terms of polynomials of
$1-u$,
\begin{eqnarray}
\phi(u)=\sum_{k=0}^N b_k(1-u)^k=\sum_{k=0}^N b_k
\left(\frac{s-m_c^2}{s-q^2}\right)^k,
\end{eqnarray}
then introduce the variable $s'$ and the spectral density is
obtained.

After straightforward calculations, we obtain the final expressions
of the double Borel transformed correlation functions  $\Pi_{ij}^P$
and $\Pi_{ij}^V$  at the level of quark-gluon degrees of freedom.
The masses of  the charmed mesons are $M_{D}=1.87\rm{GeV}$,
$M_{D_s}=1.97\rm{GeV}$, $M_{D^*}=2.010\rm{GeV}$ and
$M_{D_s^*}=2.112\rm{GeV}$.
\begin{eqnarray}
\frac{M_{D^*}}{M_{D^*}+M_{D^*_s}}\approx0.49 \, , &&
\frac{M_{D}}{M_{D}+M_{D^*}}\approx0.48 \, , \nonumber\\
\frac{M_{D_s}}{M_{D_s}+M_{D^*}}\approx0.49 \, ,&&
\frac{M_{D_s}}{M_{D_s}+M_{D_s^*}}\approx0.48 \, ,
\end{eqnarray}
 there exist overlapping working windows for the two Borel
parameters $M_1^2$ and $M_2^2$, it is convenient to take the value
$M_1^2=M_2^2$.  We introduce the threshold parameters $s_0$ and make
the simple replacement,
\begin{eqnarray}
e^{-\frac{m_c^2+u_0(1-u_0)m_{\rho,K^*,\phi}^2}{M^2}} \rightarrow
e^{-\frac{m_c^2+u_0(1-u_0)m_{\rho,K^*,\phi}^2}{M^2}
}-e^{-\frac{s^0_{\rho,K^*,\phi}}{M^2}} \nonumber
\end{eqnarray}
 to subtract the contributions from the high resonances  and
  continuum states \cite{Belyaev94}. Finally we obtain the sum rules for the strong coupling
  constants $g_{DDV}$ and $f_{D^*DV}$,
\begin{eqnarray}
&&2g_{D_iD_j V_{ij}}\frac{f_{D_i} f_{D_j}M^2_{D_i}M^2_{D_j}}{(m_c+m_i)(m_c+m_j)} \exp\left\{-\frac{M_{D_i}^2}{M^2_1}-\frac{M^2_{D_j}}{M_2^2}\right\}\nonumber\\
&=& f_{V_{ij}} m_{V_{ij}} M^2\phi_\parallel(u_0)\left\{\exp\left[-
\frac{m_c^2+u_0(1-u_0)m_{V_{ij}}^2}{M^2}
\right]-\exp\left[- \frac{s^0_{V_{ij}}}{M^2} \right] \right\} \nonumber\\
&&+\exp\left[- \frac{m_c^2+u_0(1-u_0)m_{V_{ij}}^2}{M^2}
\right]\left\{
\left[f_{V_{ij}}^\perp-f_{V_{ij}}\frac{m_i+m_j}{m_{V_{ij}}}\right]m_c
m_{V_{ij}}^2
 h_{||}^{(s)}(u_0) \right.\nonumber\\
&&\left.-\frac{f_{V_{ij}} m_{V_{ij}}^3A(u_0)}{4}  \left[1
+\frac{m_c^2}{M^2}\right] -2f_{V_{ij}} m_{V_{ij}}^3   \int_0^{u_0}
d\tau \int_0^\tau dt C(t)\left[1 +\frac{m_c^2}{M^2}\right]\right\}
\, ,
\end{eqnarray}

\begin{eqnarray}
&&4f_{D^*_iD_j V_{ij}}\frac{f_{D^*_i} f_{D_j}M_{D^*_i}M^2_{D_j}}{m_c+m_j} \exp\left\{-\frac{M_{D^*_i}^2}{M^2_1}-\frac{M^2_{D_j}}{M_2^2}\right\}\nonumber\\
&=& f_{V_{ij}}^\perp  M^2\phi_\perp(u_0)\left\{\exp\left[-
\frac{m_c^2+u_0(1-u_0)m_{V_{ij}}^2}{M^2}
\right]-\exp\left[- \frac{s^0_{V_{ij}}}{M^2} \right] \right\} \nonumber\\
&&+\exp\left[- \frac{m_c^2+u_0(1-u_0)m_{V_{ij}}^2}{M^2}
\right]\left\{
\left[f_{V_{ij}}-f_{V_{ij}}^\perp\frac{m_i+m_j}{m_{V_{ij}}}\right]
\frac{m_c m_{V_{ij}}g_{\perp}^{(a)}(u_0)}{2}  \right.\nonumber\\
&&\left.-\frac{f_{V_{ij}}^\perp m_{V_{ij}}^2A_\perp(u_0)}{4}
\left[1 +\frac{m_c^2}{M^2}\right] \right\} \, ,
\end{eqnarray}

where
\begin{eqnarray}
u_0&=&\frac{M_1^2}{M_1^2+M_2^2}\, , \nonumber \\
M^2&=&\frac{M_1^2M_2^2}{M_1^2+M_2^2} \, .
\end{eqnarray}

\section{Numerical result and discussion}
The input parameters are taken as $m_s=(0.14\pm 0.01 )\rm{GeV}$,
$m_c=(1.35\pm 0.10)\rm{GeV}$, $m_u=m_d=(0.0056\pm0.0016)\rm{GeV}$,
$f_\rho=(0.216\pm0.003)\rm{GeV}$,
$f_\rho^{\perp}=(0.165\pm0.009)\rm{GeV}$,
$f_{K^*}=(0.220\pm0.005)\rm{GeV}$,
$f_{K^*}^{\perp}=(0.185\pm0.010)\rm{GeV}$,
$f_\phi=(0.215\pm0.005)\rm{GeV}$,
$f_\phi^{\perp}=(0.186\pm0.009)\rm{GeV}$, $m_\rho=0.775\rm{GeV}$,
$m_{K^*}=0.892\rm{GeV}$, $m_\phi=1.02\rm{GeV}$, $\zeta_4=0.15\pm
0.10$, $\zeta_4^T=0.10\pm 0.05$ and $\widetilde{\zeta}_4^T=-0.10\pm
0.05$ \cite{VMLC}. The parameters in the two-particle twist-2 and
twist-3 light-cone distribution amplitudes are shown in Table.1
\cite{VMLC}.

The values of the decay constants $f_D$ and $f_{D_s}$ vary in a
large range from different approaches, for example, the potential
model, QCD sum rules and Lattice QCD, etc \cite{decayC2}.  For the
decay constant $f_D$, we take the experimental data from the CLEO
Collaboration, $f_D=(0.223\pm0.017)\rm{GeV}$  \cite{decayC}. If we
take the value $f_{D_s}=(0.274\pm0.013)\rm{GeV}$ from the CLEO
Collaboration, the $SU(3)$ breaking effect is rather large,
$\frac{f_{D_s}}{f_D}=1.23$, while most theoretical estimations
indicate  $\frac{f_{D_s}}{f_D}\approx 1.1$. In this article, we take
the value $\frac{f_{D_s}}{f_D} = 1.1$. For the decay constants
$f_{D^*}$ and $f_{D^*_s}$,  we take the
 central values  from lattice simulation \cite{decayCV},
$f_{D^*}=(0.23\pm0.02)\rm{GeV}$ and
$f_{D^*_s}=(0.25\pm0.02)\rm{GeV}$,
\begin{eqnarray}
\frac{f_{D^*_s}}{f_{D^*}}&\approx & \frac{f_{D_s}}{f_D} = 1.1 \, .
\end{eqnarray}

 The duality threshold parameters  $s_0$ are shown in Table.2, the numerical (central) values of $s_0$ are
taken from the QCD sum rules for the masses of the pseudoscalar
mesons $D^0$, $D^+$, $D_s$ and vector mesons $D^{*0}$, $D^{*+}$,
$D^*_s$ \cite{Threshold}. In this article, we take the uncertainties
for  the threshold parameters $s_0$ to be $0.5\rm{GeV}^2$ for
simplicity. The Borel parameters are chosen as $ M_1^2=M_2^2$ and
$M^2=(3-7)\rm{GeV}^2$, in those regions, the values of the strong
coupling constants $g_{DDV}$ and $f_{D^*D V}$ are rather stable.

\begin{table}[tb]
 $$
\begin{array}{|c | c  |  c | c |}
\hline\hline & {\rho} & {K^*}  &
{\phi}\\
\hline a_1^\parallel & 0  & \phantom{-}0.03(2) & 0\\
a_1^\perp & 0  & \phantom{-}0.04(3)  & 0\\
a_2^\parallel & 0.15(7)  & \phantom{-}0.11(9)  & 0.18(8)\\
a_2^\perp &  0.14(6) & \phantom{-}0.10(8)  & 0.14(7)\\\hline
\zeta_{3V}^\parallel & 0.030(10)  & \phantom{-}0.023(8)  & 0.024(8)\\
\widetilde\lambda_{3V}^\parallel & 0  & \phantom{-}0.035(15)& 0\\
\widetilde\omega_{3V}^\parallel & -0.09(3)  & -0.07(3)  & -0.045(15)\\
\kappa_{3V}^\parallel & 0 & \phantom{-}0.000(1) & 0\\
\omega_{3V}^\parallel & 0.15(5)  & \phantom{-}0.10(4) & 0.09(3)\\
\lambda_{3V}^\parallel & 0  & -0.008(4)  & 0\\
\kappa_{3V}^\perp & 0 &  \phantom{-}0.003(3)  & 0\\
\omega_{3V}^\perp & 0.55(25)  & \phantom{-}0.3(1) & 0.20(8)\\
\lambda_{3V}^\perp & 0  & -0.025(20)  & 0 \\
\hline\hline
\end{array}
$$
  \caption{The
parameters in the twist-2 and twist-3 light-cone distribution
amplitudes (taken from the last article of Ref.\cite{VMLC}).}
\end{table}

\begin{table}
\begin{center}
\begin{tabular}{|c|c|c|}
\hline\hline
      &$g_{DDV}$  & $f_{D^*DV}$ \\ \hline
$ s^0_{\rho}(\rm{GeV}^2)$     &$6.0\pm 0.5$  & $6.5\pm 0.5$ \\
$ s^0_{K^*}(\rm{GeV}^2)$     &$6.3\pm 0.5$  & $7.0\pm 0.5$ \\
$ s^0_{\phi}(\rm{GeV}^2)$     &$6.3\pm 0.5$  & $7.0\pm 0.5$ \\
      \hline  \hline
\end{tabular}
\end{center}
\caption{ Threshold parameters for the strong coupling constants
$g_{DDV}$  and $f_{D^*DV}$. }
\end{table}

In the limit of large Borel parameter $M^2$, the strong coupling
constants $g_{DDV}$ and $f_{D^*DV}$   take up the following
behaviors,
\begin{eqnarray}
g_{D_iD_jV_{ij}} &\propto&  \frac{M^2f_{V_{ij}} \phi_\parallel(u_0)}{f_{D_i}f_{D_j}}\propto  \frac{M^2f_{V_{ij}} a^\parallel_2}{f_{D_i}f_{D_j}}\, ,\nonumber\\
f_{D_i^*D_jV_{ij}}&\propto& \frac{M^2f_{V_{ij}}^\perp
\phi_\perp(u_0)}{f_{D^*_i}f_{D_j}}\propto \frac{M^2f_{V_{ij}}^\perp
a_2^\perp}{f_{D^*_i}f_{D_j}}\, .
\end{eqnarray}
It is not unexpected, the contributions  from the twist-2 light-cone
distribution amplitudes  $\phi_\parallel(u)$ and $\phi_\perp(u)$ are
greatly enhanced by the large Borel parameter $M^2$, (large)
uncertainties of the relevant parameters presented in above
equations have significant impact on the numerical results.

Taking into account all the uncertainties, finally we obtain the
numerical values for  the strong coupling constants $g_{DDV}$  and
$f_{D^*DV}$, which are shown in Figs.(1-2),
\begin{eqnarray}
  g_{DD\rho} &=&1.31 \pm 0.29 \, , \nonumber \\
g_{DD_sK^*} &=&1.61 \pm 0.32 \, , \nonumber \\
g_{D_sD_s\phi} &=&1.45 \pm 0.34 \, , \nonumber \\
f_{D^*D\rho} &=&(0.89 \pm 0.15) \rm{GeV}^{-1}\, , \nonumber \\
f_{D^*D_sK^*} &=&(1.01 \pm 0.20) \rm{GeV}^{-1}\, , \nonumber \\
f_{D^*_sD_s\phi} &=&(0.82 \pm 0.16)\rm{GeV}^{-1} \, .
\end{eqnarray}
Taking  the replacements $g_{DD\rho}\rightarrow
\frac{g_{DD\rho}}{2}$ and $f_{D^*D\rho}\rightarrow
\frac{f_{D^*D\rho}}{4}$ in Eq.(1),  we can obtain the same
definitions for the strong coupling constants in Ref.\cite{LiHuang}.
Our numerical values $g_{DD\rho} =2.62 \pm 0.58$ and $f_{D^*D\rho}
=(3.56 \pm 0.60)\rm{GeV}^{-1}$ are compatible with the predictions
$g_{DD\rho} =3.81 \pm 0.88$ and $f_{D^*D\rho} =(4.17 \pm 1.04)
\rm{GeV}^{-1}$ in Ref.\cite{LiHuang}. In Ref.\cite{LiHuang}, the
authors take much smaller values for the decay constants of the
charmed mesons than the present work. It is not unexpected that the
numerical values are different from each other, see Eq.(21).

 The average values of the strong coupling constants are about
\begin{eqnarray}
  g_{DDV} &=&1.46 \pm 0.32 \, ,\nonumber \\
f_{D^*DV} &=&(0.91 \pm 0.17) \rm{GeV}^{-1} \, .
\end{eqnarray}
The corresponding  basic parameters $\beta$ and $\lambda$ in the
heavy quark effective theory are listed in Table.3 and Table.4,
respectively. The parameter $\beta$ can be estimated with  the
vector meson dominance theory\footnote{ In this footnote, we
illustrate the estimation of the basic parameter $\beta$ with the
vector meson dominance theory.
\begin{eqnarray}
 &&
f(p^2)(p_1+p_2)_\mu
\nonumber \\
&=& \langle D_s(p_1)| \bar{s}(0)\gamma_\mu s(0) |D_s(p_2) \rangle \nonumber \\
 &=&\langle D_s(p_1)  \phi(p)| D_s(p_2) \rangle
\frac{i}{m_\phi^2-p^2}  \langle 0| \bar{s}(0)\gamma_\mu s(0)
|\phi(p)\rangle
   \nonumber\\
&=&\frac{1}{p^2-m_\phi^2} f_\phi m_\phi g_{D_sD_s\phi} \epsilon^*
\cdot (p_1+p_2)\epsilon_\mu \nonumber \\
&=&\frac{1}{p^2-m_\phi^2} f_\phi m_\phi g_{D_sD_s\phi} (p_1+p_2)_\nu
\left\{-g_{\mu\nu}+\frac{(p_1-p_2)_\mu(p_1-p_2)_\nu}{(p_1-p_2)^2}
\right\} \nonumber \\
&=&-\frac{1}{p^2-m_\phi^2} f_\phi m_\phi g_{D_sD_s\phi}
(p_1+p_2)_\mu \, .
\\
&\rightarrow& f(0)=\frac{f_\phi}{m_\phi}g_{D_sD_s\phi} \, ,
\nonumber
\end{eqnarray}
 Take the normalization condition $f(0)=1$,
\begin{eqnarray}
&\rightarrow& g_{D_sD_s\phi}=\frac{m_\phi}{f_\phi}\, ,\nonumber \\
&\rightarrow& \frac{\beta g_V}{\sqrt{2}}=\frac{m_\phi}{f_\phi} \, ,
\,\,\, \rm{see \, \, Ref.}\cite{VMD03} \, .
\end{eqnarray}
If we take into account the contribution from the $2^3S_1$ state
$\phi(1680)$, the expression would  be
\begin{eqnarray}
1=\frac{f_\phi}{m_\phi}g_{D_sD_s\phi}+\frac{f_{\phi(1680)}}{m_{\phi(1680)}}g_{D_sD_s\phi(1680)}
\, .
\end{eqnarray}
If the value of the $g_{D_sD_s\phi(1680)}$ is positive,  much
smaller value of the $\beta$ can be obtained. For example, with the
assumption $g_{D_sD_s\phi(1680)}=g_{D_sD_s\phi}$ and
$f_{\phi(1680)}=f_{\phi}$, we can obtain  $\frac{\beta
g_V}{\sqrt{2}}=\frac{m_\phi m_{\phi(1680)}}{(m_\phi+
m_{\phi(1680)})f_\phi}=\frac{0.62 m_\phi  }{ f_\phi} $, the value of
the $\beta$ listed in Table.3 would be $\beta\approx 0.62\times
0.9\approx 0.56$, our prediction is still much smaller.
 }, which is presented in Table.3.
The basic parameter $\lambda$  relates to  the form-factor $V(q^2)$
of the hadronic transitions $\langle V\mid \bar{q}
\gamma_\mu(1-\gamma_5)b \mid B \rangle $ and $\langle V\mid
\bar{q}\sigma_{\mu\nu}(1+\gamma_5)b \mid B \rangle $,
  which can be calculated with the
light-cone sum rules  and lattice QCD. With assumption that  the
form-factor $V(q^2)$ at $q^2= q^2_{max}= (M_B-M_V)^2$ is dominated
by the nearest low-lying vector meson pole, we can obtain the values
of the $\lambda$ \cite{VMD03,CasalbuoniLammbda}, which are presented
in Table.4.  From the Tables.3-4, we
 can see that our numerical values are much smaller.

 One possibility for the large  discrepancies  maybe that the vector meson dominance theory overestimates
  the values of the $\beta g_V$ and
 $\lambda g_V$,
 the other possibility maybe
 the  shortcomings of the light-cone QCD sum rules.
We can borrow some idea from the
  strong coupling constant $g_{D^*D\pi}$, the central value
($g_{D^*D\pi}=12.5$ or $g_{D^*D\pi}=10.5$ with the radiative
corrections are included in) from the light-cone QCD sum rules is
too small to take into account
  the value ($g_{D^*D\pi}=17.9$) from the
experimental data \cite{Belyaev94,Khodjamirian99,Becirevic03}. It
has been noted that the simple quark-hadron duality ansatz which
works in the one-variable dispersion relation might be too crude for
the double dispersion relation \cite{KhodjamirianConf}. As in
Ref.\cite{Becirevic03},  we can postpone the threshold parameters
$s_0$ to larger values to include the contributions from  a radial
excitation ($D'$ or ${D^{*}}'$) to the hadronic spectral densities,
with additional assumption for the values of the $g_{D'DV}$,
$f_{{D^*}'DV}$ and $f_{D^*D'V}$,  we can improve the values of the
$g_{DDV}$ and $f_{D^*DV}$, and smear the discrepancies  between our
values and the predictions with the vector meson dominance theory.
It is somewhat of fine-tuning.

\begin{table}
\begin{center}
\begin{tabular}{|c|c|}
\hline\hline
      $\beta$  &Reference  \\ \hline
            $0.9$& \cite{VMD03} \\ \hline
   $0.36\pm0.08$  &  This work
\\ \hline  \hline
\end{tabular}
\end{center}
\caption{ Numerical values of the  parameter $\beta$. }
\end{table}

\begin{table}
\begin{center}
\begin{tabular}{|c|c|}
\hline\hline
      $|\lambda|(\rm{GeV}^{-1})$  &Reference  \\ \hline
      $0.56$& \cite{VMD03} \\ \hline
      $0.63\pm0.17$ &\cite{CasalbuoniLammbda}\\      \hline
          $0.22\pm0.04$  &  This work
\\ \hline  \hline
\end{tabular}
\end{center}
\caption{ Numerical values of the  parameter $\lambda$. }
\end{table}

 Naively, we can expect that  smaller values of the strong
coupling constants lead to smaller final-state re-scattering effects
in the hadronic $B$ decays. For example, the contributions from the
re-scattering mechanism for the decay
\begin{eqnarray}
B\rightarrow D^* \rho \rightarrow D\pi \nonumber
\end{eqnarray}
can occur  through exchange of $D^*$ (or $D$) in the $t$ channel for
the sub-precess $D^* \rho \rightarrow D\pi$ \cite{CHHY}. The
amplitude of the re-scattering Feynman diagrams  is proportional to
\begin{eqnarray}
C_1g_{D^*D^*\pi}f_{D^*D\rho}+C_2g_{D^*D\pi}g_{DD\rho}  \, ,
\end{eqnarray}
where the $C_i$ are some  coefficients.

\begin{figure}
\centering
  \includegraphics[totalheight=6cm,width=7cm]{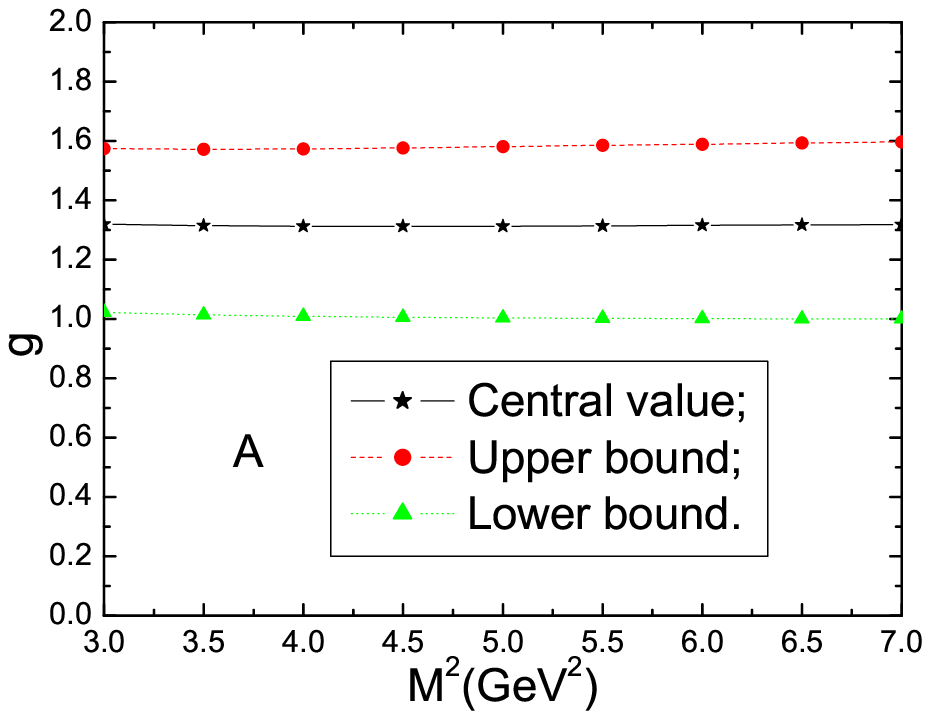}
  \includegraphics[totalheight=6cm,width=7cm]{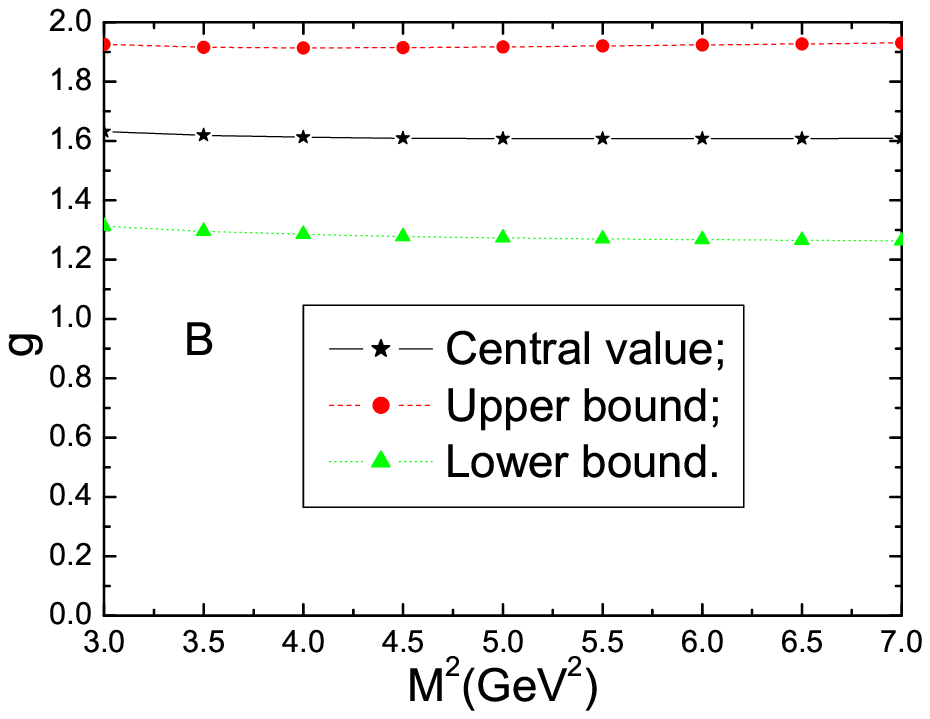}
\includegraphics[totalheight=6cm,width=7cm]{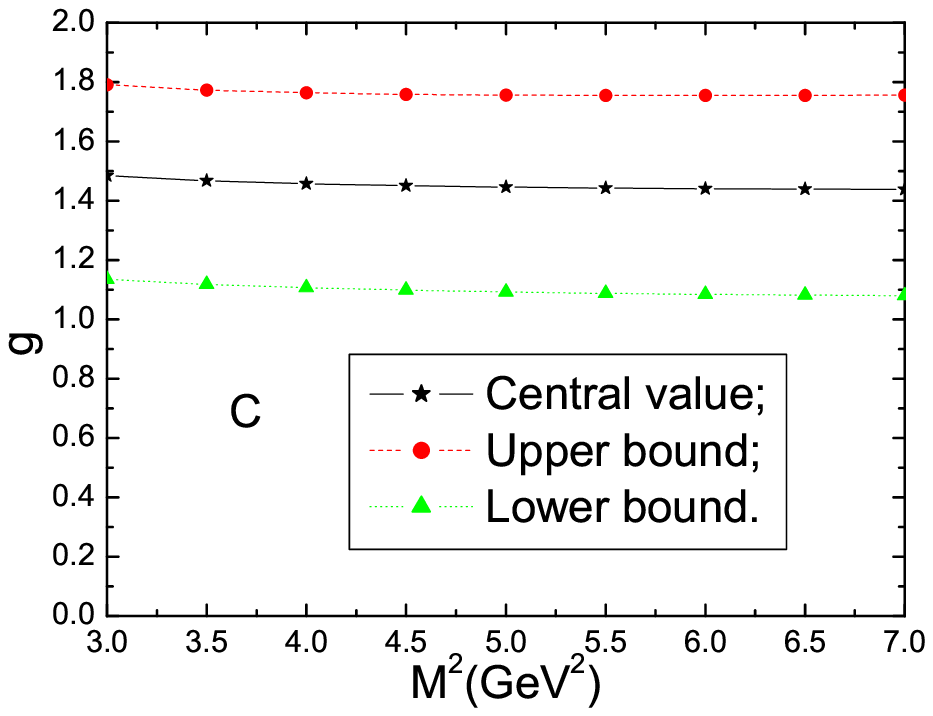}
         \caption{  $g_{DD\rho}$(A), $g_{DD_sK^*}$(B) and $g_{D_sD_s\phi}$(C)  with the Borel parameter $M^2$. }
\end{figure}

\begin{figure}
\centering
  \includegraphics[totalheight=6cm,width=7cm]{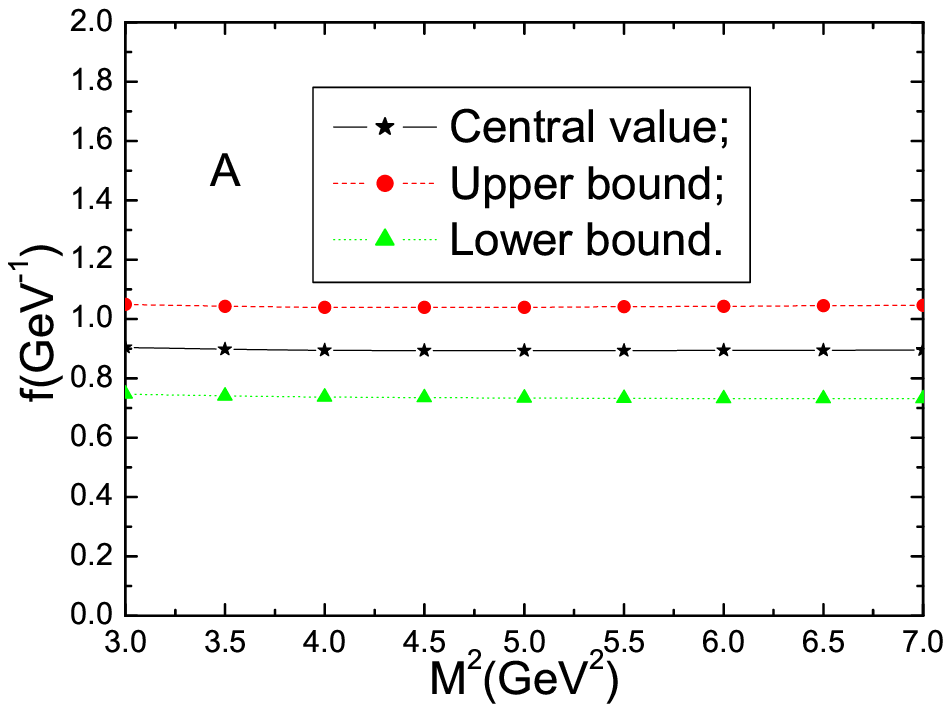}
  \includegraphics[totalheight=6cm,width=7cm]{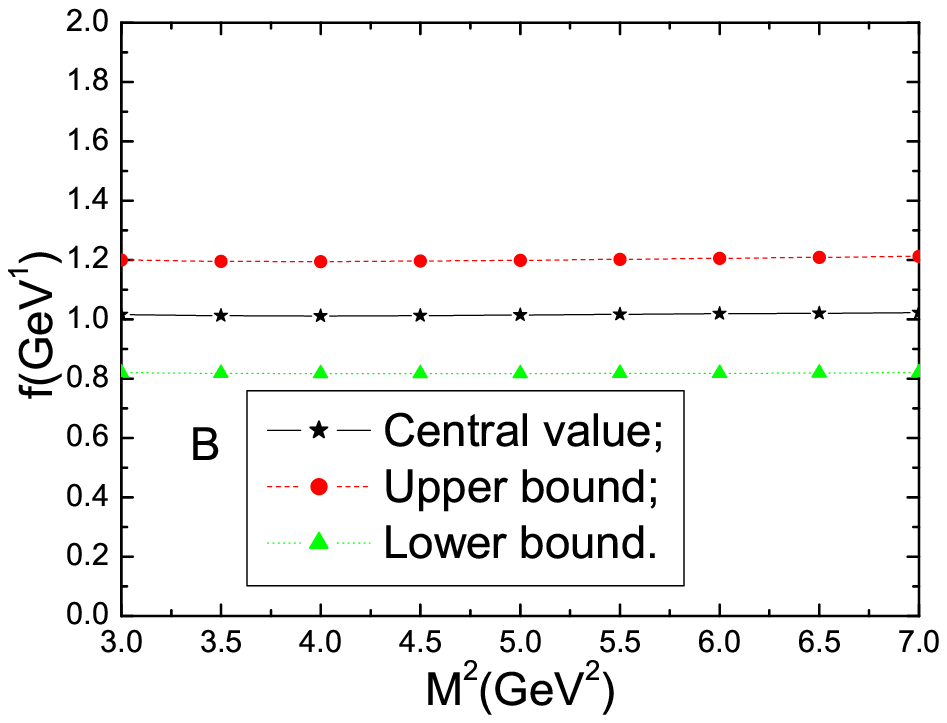}
\includegraphics[totalheight=6cm,width=7cm]{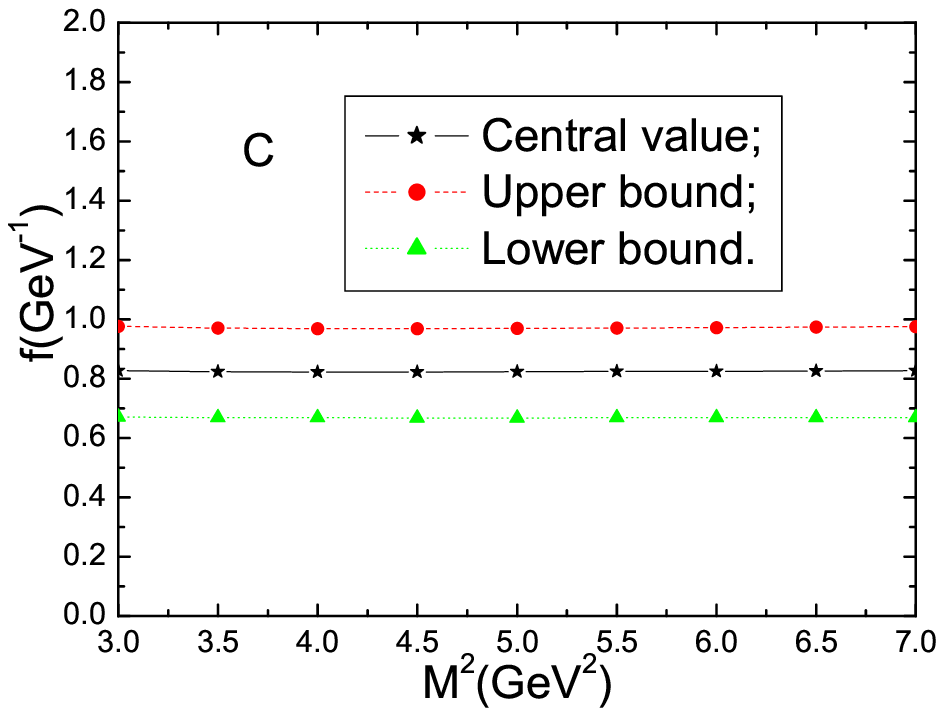}
         \caption{  $f_{D^*D\rho}$(A), $f_{D^*D_sK^*}$(B) and $f_{D^*_sD_s\phi}$(C)  with the Borel parameter $M^2$. }
\end{figure}

\section{Conclusion}
In this article, we study the vertices $DDV$ and $D^*DV$ with
  the light-cone QCD sum rules.  The  strong coupling
constants $g_{DDV}$ and $f_{D^*DV}$ play an important role in
understanding the final-state re-scattering effects  in the hadronic
$B$ decays. They are related to the basic parameters $\beta$ and
$\lambda$ in the heavy quark effective Lagrangian, the numerical
values  are much smaller than the existing estimations based on the
assumption of vector mesons  dominance. If the predictions from the
light-cone QCD sum rules are robust, the final-state re-scattering
effects maybe overestimated in the hadronic $B$ decays.
 \section*{Appendix}
The light-cone distribution amplitudes of the $K^*$ meson are
defined
 by
\begin{eqnarray}
\langle 0| {\bar u} (0) \gamma_\mu s(x) |K^*(p)\rangle& =& p_\mu
f_{K^*} m_{K^*} \frac{\epsilon \cdot x}{p \cdot x} \int_0^1 du e^{-i
u\, p\cdot x} \left\{\phi_{\parallel}(u)+\frac{m_{K^*}^2x^2}{16}
A(u)\right\}\nonumber\\
&&+\left[ \epsilon_\mu-p_\mu \frac{\epsilon \cdot x}{p \cdot x}
\right]f_{K^*} m_{K^*}
\int_0^1 du  e^{-i u \,p \cdot x} g_{\perp}^{(v)}(u)  \nonumber\\
&&-\frac{1}{2}x_\mu \frac{\epsilon \cdot x}{(p \cdot x)^2} f_{K^*} m_{K^*}^3 \int_0^1 du e^{-iu\,p \cdot x}C(u) \, ,\nonumber\\
 \langle 0| {\bar u} (0)  s(x) |{K^*}(p)\rangle  &=& \frac{i}{
2}\left[f_{K^*}^\perp-f_{K^*}
\frac{m_u+m_s}{m_{K^*}}\right]m_{K^*}^2\epsilon \cdot x  \int_0^1 du
e^{-i u \,p \cdot x} h_{\parallel}^{(s)}(u)  \,
,  \nonumber\\
\langle 0| {\bar u} (0) \sigma_{\mu \nu}  s(x) |{K^*}(p)\rangle
&=&i[\epsilon_\mu p_\nu-\epsilon_\nu p_\mu] f_{K^*}^\perp \int_0^1
du e^{-i u\, p \cdot x}
\left\{\phi_{\perp}(u)+\frac{m_{K^*}^2x^2}{16}
A_{\perp}(u) \right\}   \nonumber\\
&&+i[p_\mu x_\nu-p_\nu x_\mu] f_{K^*}^\perp m_{K^*}^2\frac{\epsilon
\cdot x}{(p \cdot x)^2} \int_0^1 du e^{-i u\, p \cdot x}
 B_{\perp}(u)   \nonumber\\
 &&+i\frac{1}{2}[\epsilon_\mu x_\nu-\epsilon_\nu x_\mu] f_{K^*}^\perp m_{K^*}^2\frac{1}{p
\cdot x} \int_0^1 du e^{-i u \,p \cdot x}
 C_{\perp}(u)   \, ,\nonumber\\
\langle 0| {\bar u} (0) \gamma_\mu \gamma_5 s(x) |{K^*}(p)\rangle
&=& -\frac{1}{ 4}\left[f_{K^*}-f_{K^*}^\perp
\frac{m_u+m_s}{m_{K^*}}\right]m_{K^*}
\epsilon_{\mu\nu\alpha\beta}\epsilon^\nu p^\alpha x^\beta
\nonumber\\
&&\int_0^1 du e^{-i u \,p \cdot x} g_{\perp}^{(a)}(u)  \, .
\end{eqnarray}
The  light-cone distribution amplitudes of the $K^*$ meson are
parameterized as
\begin{eqnarray}
\phi_{\parallel}(u,\mu)&=&6u(1-u)
\left\{1+a_1^{\parallel}3\xi+a_2^{\parallel}
\frac{3}{2}(5\xi^2-1) \right\}\, , \nonumber\\
\phi_{\perp}(u,\mu)&=&6u(1-u) \left\{1+a_1^{\perp}3\xi+a_2^{\perp}
\frac{3}{2}(5\xi^2-1) \right\}\, , \nonumber\\
g_{\perp}^{(v)}(u,\mu)&=&\frac{3}{4}(1+\xi^2)+a_1^{\parallel}\frac{3}{2}\xi^3+\left\{ \frac{3}{7}a_2^{\parallel}+ 5\zeta_3^{\parallel}\right\}(3\xi^2-1)\nonumber \\
&&+\left\{ 5\kappa_3^{\parallel}-\frac{15}{16}\lambda_3^{\parallel}+\frac{15}{8}\widetilde{\lambda}_3^{\parallel}\right\}\xi(5\xi^2-3)\nonumber\\
&&+\left\{\frac{9}{112}a_2^{\parallel}+\frac{15}{32}\omega_3^{\parallel}
-\frac{15}{64}\widetilde{\omega}_3^{\parallel}\right\}(3-30\xi^2+35\xi^4)\,
,  \nonumber\\
g_{\perp}^{(a)}(u,\mu)&=& 6u\bar{u} \left\{1+\left(\frac{1}{3}a_1^{\parallel}+\frac{20}{9}\kappa_3^{\parallel} \right) C_1^{\frac{3}{2}}(\xi) + \right. \nonumber \\
&&\left.\left(\frac{1}{6}a_2^{\parallel}
+\frac{10}{9}\zeta_3^{\parallel}+\frac{5}{12}\omega_3^{\parallel}-\frac{5}{24}\widetilde{\omega}_3^{\parallel}\right)C_2^{\frac{3}{2}}(\xi)
+\left(
\frac{1}{4}\widetilde{\lambda}_3^\parallel-\frac{1}{8}\lambda_3^\parallel
\right)C_3^{\frac{3}{2}}(\xi) \right\} \nonumber\\
 h_{\parallel}^{(s)}(u,\mu)&=&6u\bar{u}
\left\{1+\left(\frac{a_1^\perp}{3}+\frac{5}{3}\kappa^{\perp}_3
\right)C_1^{\frac{3}{2}}(\xi)+\left(\frac{a_2^\perp}{6}+\frac{5}{18}\omega^{\perp}_3
\right)C_2^{\frac{3}{2}}(\xi)-\frac{1}{20}\lambda_3^{\perp}C_3^{\frac{3}{2}}(\xi) \right\}\, , \nonumber\\
h_{\parallel}^{(t)}(u,\mu)&=&3\xi^2+\frac{3}{2}a_1^{\perp}\xi(3\xi^2-1)+\frac{3}{2}a_2^{\perp}\xi^2(5\xi^2-3)+\frac{5}{8}
\omega_3^{\perp}(3-30\xi^2+35\xi^4)   \nonumber \\
&&+\left( \frac{15}{2}\kappa_3^{\perp}
-\frac{3}{4}\lambda_3^{\perp}\right)\xi(5\xi^2-3) \, ,\nonumber\\
g_3(u,\mu)&=&1+\left\{ -1-\frac{2}{7}a_2^{\parallel}+\frac{40}{3}\zeta_3^{\parallel} -\frac{20}{3}\zeta_4\right\}C_2^{\frac{1}{2}}(\xi)\nonumber \\
&&+\left\{-\frac{27}{28}a_2^{\parallel}
+\frac{5}{4}\zeta_3^{\parallel}
-\frac{15}{16}\widetilde{\omega}_3^{\parallel} -\frac{15}{8}\omega_3^{\parallel}\right\}C_4^{\frac{1}{2}}(\xi)\, ,  \nonumber \\
h_3(u,\mu)&=&1+\left\{ -1+\frac{3}{7}a_2^{\perp}-10(\zeta_4^T+\widetilde{\zeta}_4^T)\right\}C_2^{\frac{1}{2}}(\xi)+\left\{-\frac{3}{7}a_2^{\perp} -\frac{5}{4}\omega_3^{\perp}\right\}C_4^{\frac{1}{2}}(\xi)\, ,  \nonumber \\
A(u,\mu)&=&30u^2\bar{u}^2\left\{\frac{4}{5}
+\frac{4}{105}a_2^{\parallel}+\frac{8}{9}\zeta_3^{\parallel}+\frac{20}{9}\zeta_4\right\}\,
, \nonumber \\
A_{\perp}(u,\mu)&=&30u^2\bar{u}^2\left\{\frac{2}{5}
+\frac{4}{35}a_2^{\perp}+\frac{4}{3}\zeta_4^T-\frac{8}{3}\widetilde{\zeta}_4^T\right\}\,
, \nonumber \\
C(u,\mu)&=&g_3(u,\mu)+\phi_{\parallel}(u,\mu)-2g_{\perp}^{(v)}(u,\mu)
\, ,\nonumber \\
B_{\perp}(u,\mu)&=&h_{\parallel}^{(t)}(u,\mu)-\frac{1}{2}\phi_{\perp}(u,\mu)-\frac{1}{2}h_3(u,\mu)\, ,\nonumber \\
C_{\perp}(u,\mu)&=&h_3(u,\mu)-\phi_{\perp}(u,\mu) \, ,
\end{eqnarray}
where   $\xi=2u-1$, and  $C_2^{\frac{1}{2}}(\xi)$, $
C_4^{\frac{1}{2}}(\xi)$,
 $ C_1^{\frac{3}{2}}(\xi)$, $ C_2^{\frac{3}{2}}(\xi)$,  $
 C_3^{\frac{3}{2}}(\xi)$
   are Gegenbauer polynomials. The corresponding light-cone
 distribution amplitudes for the $\rho$ and $\phi$ mesons can be obtained with a
 simple replacement of the nonperturbative parameters.
\section*{Acknowledgments}
This  work is supported by National Natural Science Foundation,
Grant Number 10405009,  and Key Program Foundation of NCEPU.

\end{document}